\documentclass[a4paper,11pt]{article}
\usepackage{pos}
\usepackage{slashed, macros}
\graphicspath{{figures/}}

\title{Exploring interpolating momentum schemes }

\author*[a,b]{N. Garron}
\author[b]{C. Cahill}
\author[b]{M. Gorbahn}
\author[b]{J. A. Gracey} 
\author[b]{P. E. L. Rakow}

\affiliation[a]{School of Mathematics, Computer Science and Engineering, 
 Liverpool Hope University, Hope Park, Liverpool L16 9JD, UK}

\affiliation[b]{Theoretical Physics Division, Department of Mathematical Sciences,
  University of Liverpool, Liverpool L69 3BX, UK}

\emailAdd{garronn@hope.ac.uk}

\abstract{
  We compute the renormalisation factors of the quark mass and wave function 
  using IMOM (Interpolating MOMenta) schemes. The framework is the Rome-Southampton
  non-renormalisation method, but the momentum transfer in the quark bilinears
  is not restricted to zero or to the symmetric point. We study the scale
  dependence, infrared contamination and lattice artefacts for different
  values of this momentum transfer and for two different kinds of projectors.
  For the numerical simulations, we use data generated by the RBC-UKQCD
  collaborations, with $N_f = 2+1$ flavours of Domain-Wall fermions,
  and inverse lattice spacing of $1.79 $ and $2.38$ GeV. 
  }
\FullConference{%
 The 38th International Symposium on Lattice Field Theory, LATTICE2021
  26th-30th July, 2021
  Zoom/Gather@Massachusetts Institute of Technology
}

\begin{document}
\maketitle

\section{Kinematics}

\begin{figure}[t]
  \begin{center}
    \begin{tabular}{cc}
      \includegraphics[scale=1]{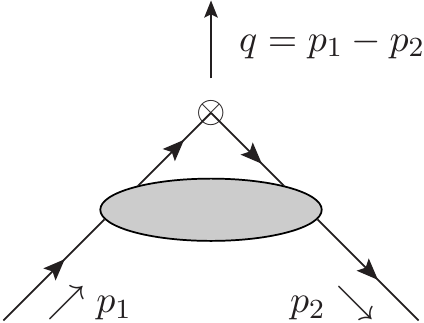} \qquad & \qquad
      \includegraphics[scale=0.8]{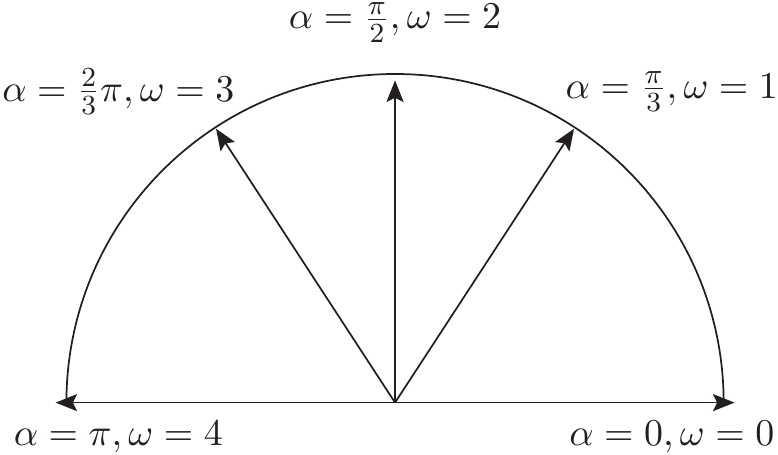}
    \end{tabular}
    \end{center}
    \caption{
      Left: a quark bilinear with incoming momentum $p_1$, outgoing momentum $p_2$ and 
      momentum transfer $q=p_1-p_2$. 
      Right:
      relationship between $\omega$ and the angle $\alpha$ between the incoming
      and outgoing momenta.}
\label{fig:bil_kin}
\end{figure}

In the framework of the Rome-Southampton method~\cite{Martinelli:1994ty},
one imposes a set of renormalisation conditions on composite operator Green's functions
computed non-perturbatively on the lattice. We consider here a generic flavour
non-singlet quark bilinear $O_\Gamma = \bar \psi_i \Gamma \psi_j$, where $i\ne j$
and $\Gamma$ is a Dirac matrix. We suppress the flavour indices $i$ and $j$ for simplicity. 
Traditionally the momentum transfer is chosen is to be either zero
or such that $p_1^2-p_2^2 = (p_1-p_2)^2$, where $p_1$ and $p_2$ are the
incoming and outgoing momenta, respectively (see Fig.~\ref{fig:bil_kin}).
The former is known to lead to exceptional kinematics and therefore potentially large
unwanted  infrared contributions; the latter is referred to as the {\em symmetric} point
and defines a so-called RI/SMOM scheme~\cite{Aoki:2007xm,Sturm:2009kb}.
The main purpose of the  RI/SMOM kinematics is to suppress the unwanted low-energy
contributions. Here we want to generalise this choice of kinematics.
As usual, the renormalisation scale is called $\mu$, 
but we define an additional parameter $\omega$ such that
\begin{eqnarray}
  (p_1-p_2)^2 &=& \omega \mu^2 \;, \\
  \mu^2 &=& p_1^2 = p_2^2 \;.
\end{eqnarray}
It follows that $\omega=0$ corresponds to zero-momentum transfer and $\omega=1$
corresponds to the RI/SMOM kinematics. Although it makes sense to fix
$\omega$ to either of these values in order to be left with only one scale in the game,
in general the parameter 
$\omega$ can take any value between $0$ and $4$. One can define an angle $\alpha$
between $p_1$ and $p_2$ and we find that $\omega = 2 (1-\cos \alpha)$,
as illustrated in Fig.~\ref{fig:bil_kin}.
It is clear that the extreme values of $\omega$ where $p_1$ and $p_2$ are parallel
or anti-parallel can lead to collinear singularities.
Letting $\omega$ vary as a free parameter defines the RI/IMOM schemes (we will
now drop the ``RI'' to ease the notations). The interested reader can find more details
in~\cite{Garron:2021mfn}. \\

\section{Definitions}
\subsection{Z-factors}
We study $Z_m$ and $Z_q$, the renormalisation factors of the quark mass and wave function,
respectively. The are defined in the chiral limit ($m$ represents the quark mass) through
\begin{eqnarray}
\label{eq:Zq}
Z_q^{(X)}(\mu,\omega) &=&  Z_V \lim_{m\to0} \left[\Lambda_V^{(X)}\right]_{\text{IMOM}} \;,\\
\label{eq:Zm}
Z_m^{(X)}(\mu,\omega)  &=& \frac{1}{Z_V}  \lim_{m\to0} \left[\frac{\Lambda_S}{\Lambda_V^{(X)}}\right]_{\text{IMOM}}   \;.
\end{eqnarray}
On the right-hand-side of Eqs.~(\ref{eq:Zq}) and (\ref{eq:Zm}), $\Lambda_{S,P}$ represent
the amputated and projected vertex functions computed on Landau-gauge fixed configurations,
at finite quark mass $m=Z_m m_{bare}$ (we take all quark masses to be same for simplicity).
The values of $Z_V$ are known from previous work~\cite{RBC:2010qam}.
The choice of projector is denoted by $X\in (\gamma_\mu,\qslash)$,
more explicitly:
\begin{eqnarray}
\label{eq:LambdaS}
\Lambda_S &=& \frac{1}{12} \text{Tr}[ \Pi_{S} ]\;, \\
\label{eq:gammamuProj}
\Lambda_V^{(\gamma_\mu)} &=& \frac{1}{48} \text{Tr}[ \gamma_\mu \Pi_{V^{\mu}} ]\;, \\
\label{eq:qslashProj}
\Lambda_V^{(\slashed{q})} &=& \frac{q^{\mu}}{12 q^2} \text{Tr}[ \slashed{q} \Pi_{V^{\mu}}\, ]\;,
\end{eqnarray}
where $\Pi_\Gamma, \Gamma=S, {V^{\mu}}$ represents the amputated vertex function:
\be
\Pi_\Gamma =
\langle G^{-1}(-p_2) \rangle V_\Gamma(p_2,p_1) \langle G^{-1}(p_1)\rangle \;,
\ee
and  
\bea
\label{Vgamma}
V_\Gamma(p_2,p_1) &=& \la \psi(p_2)  O_\Gamma \bar \psi (p_1)\ra \;, \\ 
&=& \sum_x \la G_x(-p_2) \Gamma G_x(p_1) \ra \;, \\
G(p) &=& \sum_x G_x(p) \;. 
\eea
Finally, within our conventions, $G_x(p)$ represents an incoming quark propagator
with momentum $p$, where the Fourier transform is computed at space-time point $x$, explicitly:
\be
G_x(p) = \sum_y D^{-1}(x,y)e^{ip.(y-x)}. \label{npr4}
\ee
In order to assess some systematic errors, we also implement the vertex function
for $\Lambda_A$ and $\Lambda_P$. They are defined exactly in the same way,
with $V \longrightarrow A$ and $S\longrightarrow P$ in the previous equations

\subsection{Running}
We compute the non-perturbative scale evolution of $Z_y, y\in (m,q)$,
we define $\Sigma_{m}$ as:
\be
\label{eq:sigma_def}
\Sigma^{(X)}_{y}(a,\mu,\mu_0,\omega, \omega_0) = \lim_{m\rightarrow 0}\frac{Z^{(X)}_y(a,\mu,\omega)}{Z^{(X)}_y(a,\mu_0,\omega_0)}\;,
\ee
where as above $X$ can be either $\gamma_\mu$ or $\slashed{q}$. 
We take the continuum limit :
\be
\sigma^{(X)}_{y}(\mu,\mu_0,\omega, \omega_0) = \lim_{a^2 \rightarrow 0} \Sigma^{(X)}_{y}(a,\mu,\mu_0,\omega, \omega_0) \;.
\ee
We also compute this running in perturbation theory at Next-to-Next-to-Leading Order (NNLO).
We note that for $Z_m$, the corresponding anomalous dimensions have been recently computed
in~\cite{Bednyakov:2020} and~\cite{Kniehl:2020} at N$^3$LO in the case $\omega=1$.
In $\MSbar$, they can be found in~\cite{Chetyrkin:1999pq}, together with the one
of the quark wave function for the $\qslash$-projector.

\section{Results}

\begin{figure}[t]
  \bc
  \includegraphics[width=0.8\textwidth]{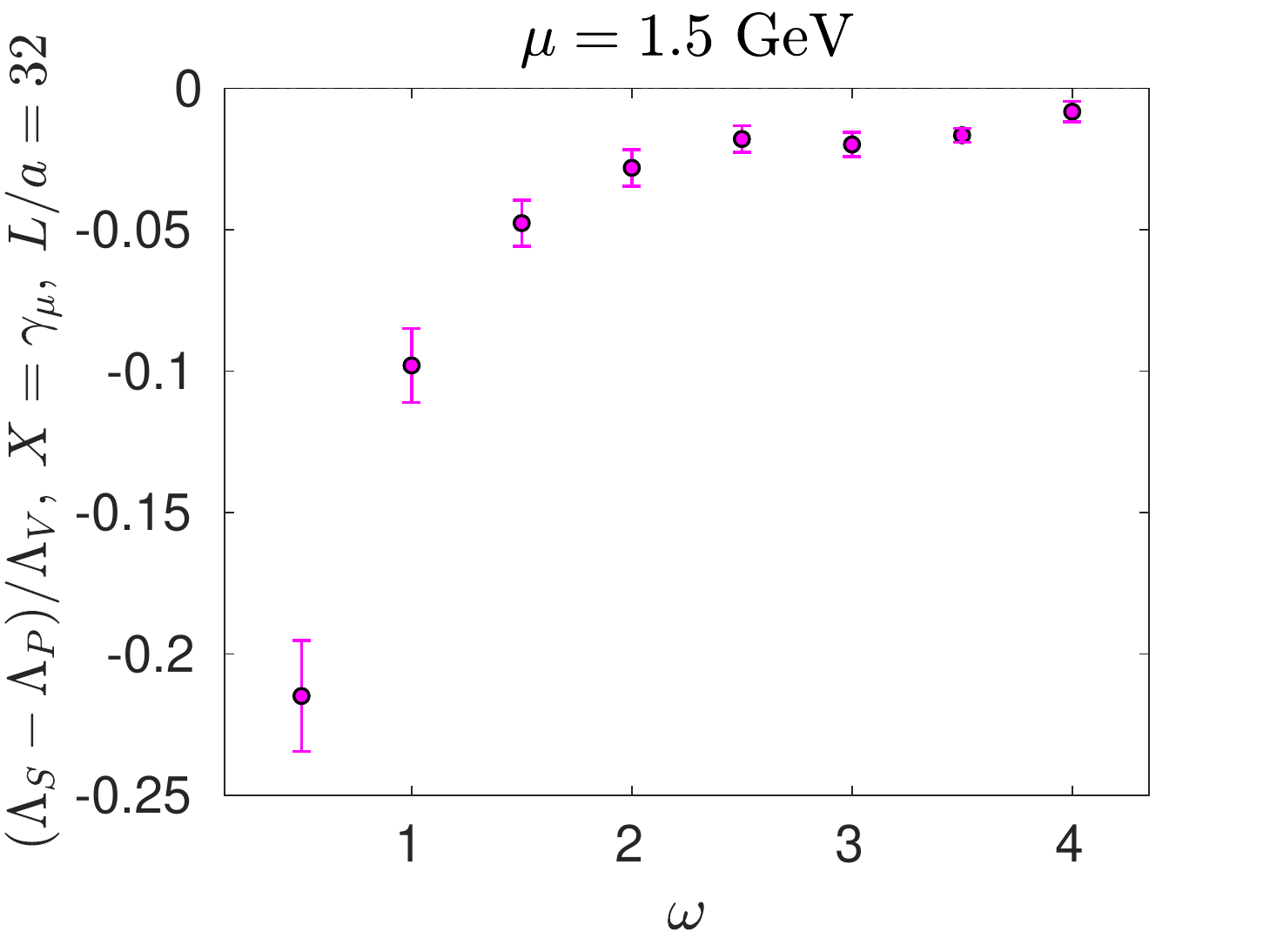}
   \ec
   \caption{As a measure of chiral symmetry breaking effects we show
     $(\Lambda_S-\Lambda_P) / \Lambda_V$ for the ${\gamma_\mu}$-projector,
     as a function of $\omega$, for a fixed value of $\mu$.
     The unwanted low energy contributions decrease quickly as $\omega$
     increases.}
\label{fig:LSmP_vs_w}
\end{figure}

As it is often the case for a NPR  study, the choice of the lattice discretisation
is of crucial importance. The good chiral-flavour properties of the Domain-Wall fermions
are essential to disentangle physical infrared contributions from
artefacts due to the choice of fermionic action. 
In absence of chiral symmetry breaking, we should find  $\Lambda_S = \Lambda_P$.
In Fig.~\ref{fig:LSmP_vs_w}, we show $(\Lambda_S - \Lambda_P)/\Lambda_V$
as a function of $\omega$, for $\mu=1.5$ GeV
(we divide by $\Lambda_V$ to cancel the quark wave function renormalisation factor).
We find that this quantity is much smaller for $\omega\ge 2$ than for $\omega=1$:
$\sim 0.03$ vs. $\sim 0.10$ .
This could be important for four-quark operators such as $(S-P)\times (S-P)$
and $(S-P)\times (S+P)$ which can also mix due to chiral symmetry breaking effects.

\begin{figure}[htb]
  \bc
  \begin{tabular}{cc}
    \includegraphics[width=0.5\textwidth]{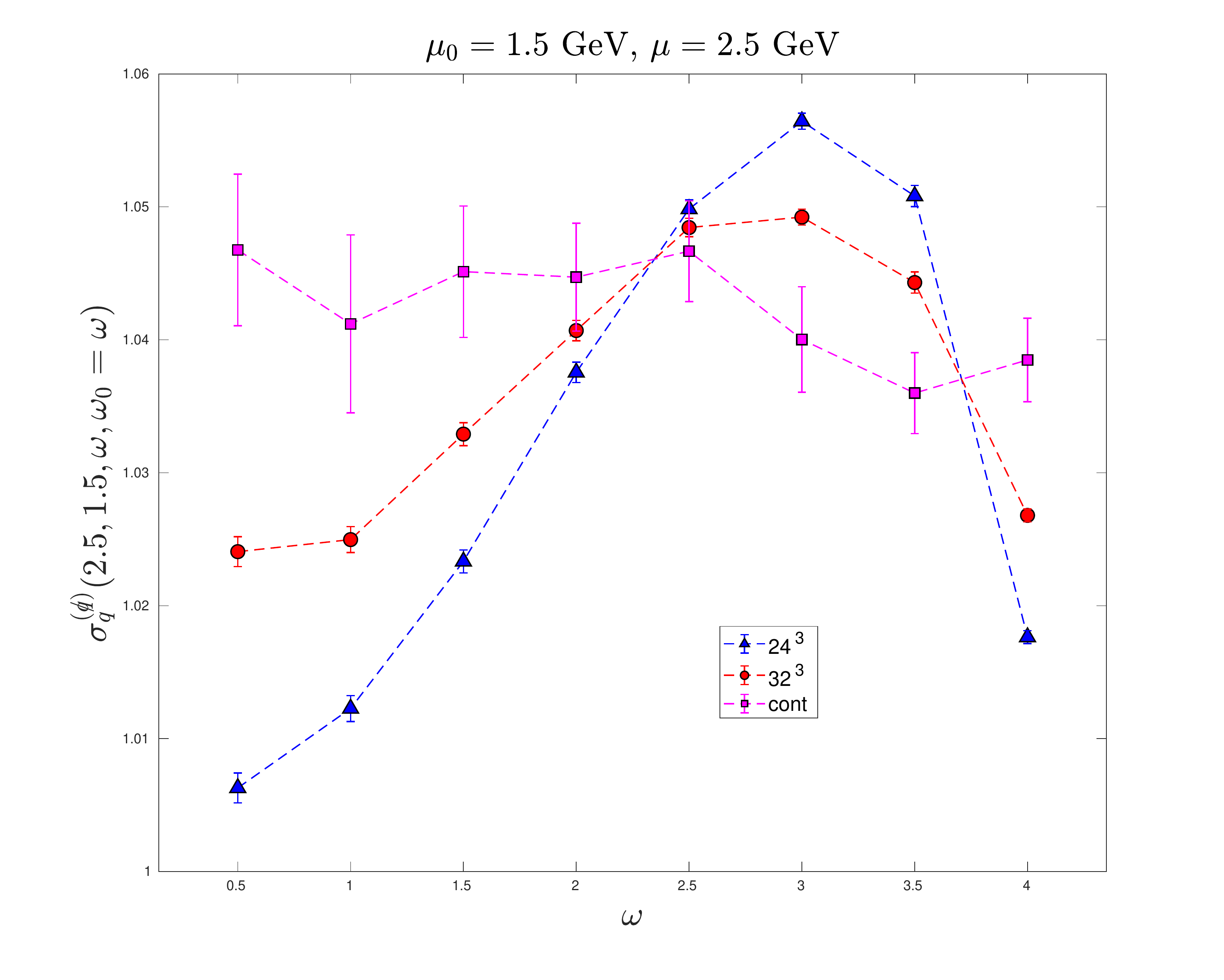} &
    \includegraphics[width=0.5\textwidth]{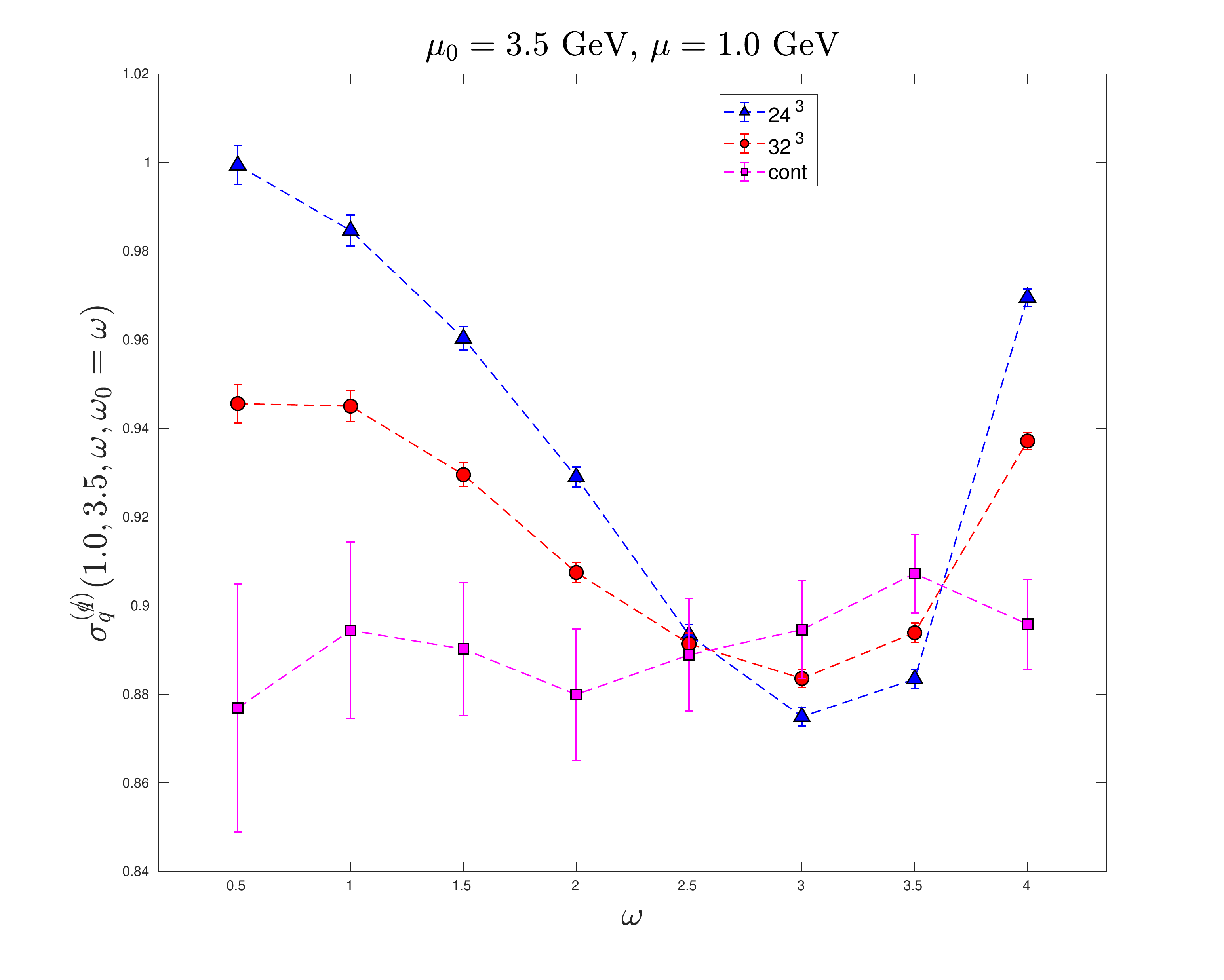} \\
    \end{tabular}
\ec
\caption{
  Example of continuum extrapolations for  $\sigma_q^{(\qslash)}(\mu,\mu_0, \omega, \omega_0) $.
}
\label{fig:cl_sigma_q_vs_w}
\end{figure}

In Fig.\ref{fig:cl_sigma_q_vs_w} we show the non-perturbative scale evolution
for $Z_q^{(\qslash)}$ at finite lattice spacing and in the continuum, for different
values of $\omega=\omega_0$. We expect this quantity to be $\omega$-independent
due to the vector Ward-Takahashi identity. Although after continuum extrapolation
this quantity is indeed $\omega$-independent (to a good approximation),
this is clearly not the case at finite lattice spacing. Using this
quantity as a measure of the discretisation effects,
Fig.\ref{fig:cl_sigma_q_vs_w} suggests that the region $\omega\sim 2.0-2.5$
is less affected by lattice artefacts (for this quantity).

\begin{figure}[t]
\bc
\includegraphics[scale=.82]{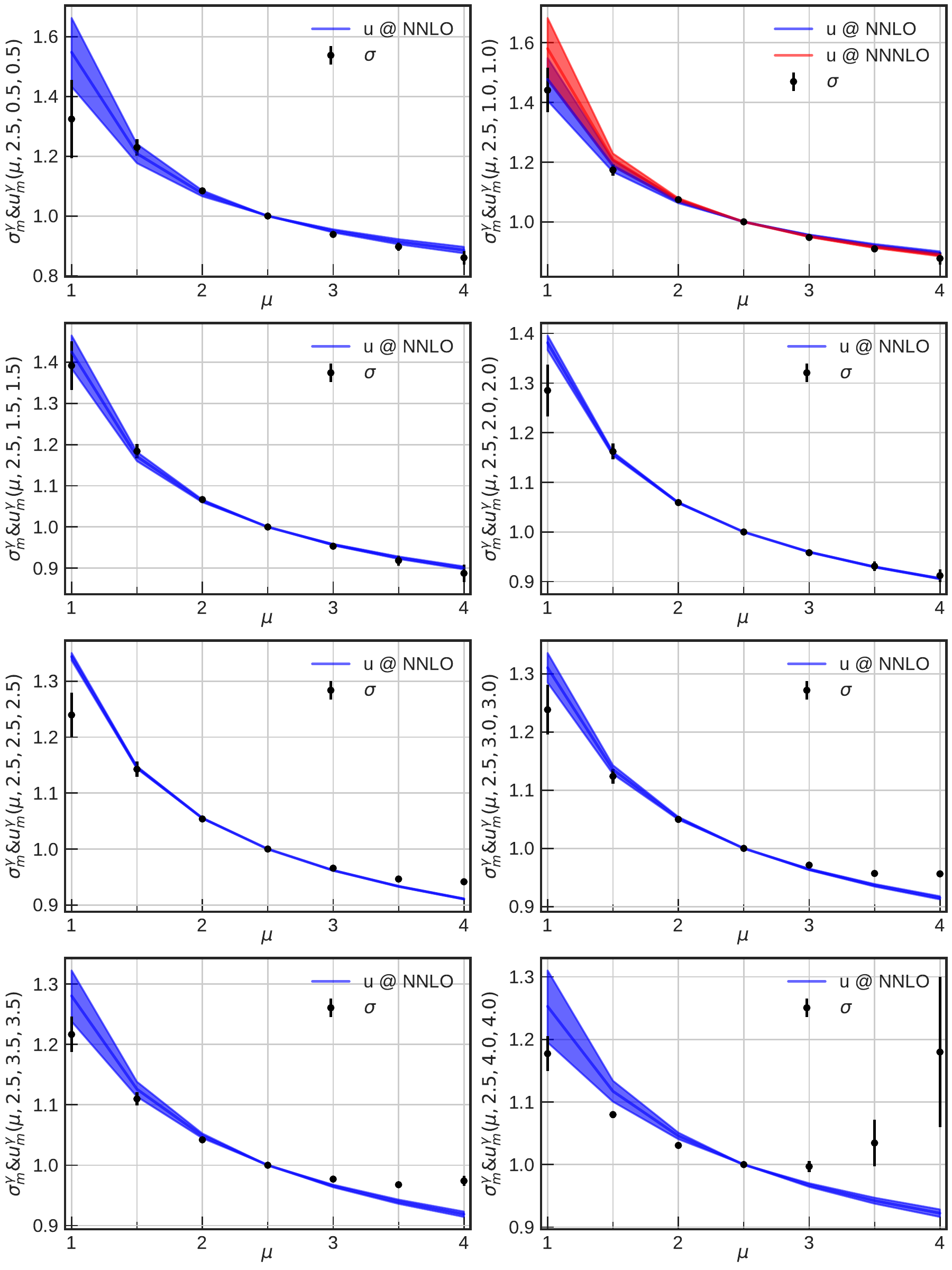}
\ec
\caption{Comparison of the non-perturbative and perturbative running
  for $Z_m^{(\gamma_\mu)}$. 
Note that for $\omega=1$ the perturbative running is known at N$^3$LO.}
\label{fig:Compare_m_gamma_mu0_25}
\end{figure}
We show the running of the quark mass in Fig.~\ref{fig:Compare_m_gamma_mu0_25}
for the $\gamma_\mu$-scheme: both the non-perturbative 
scale evolution $\sigma^{(\gamma_\mu)}_{m}(\mu,\mu_0,\omega, \omega_0)$
and the perturbative prediction $u^{(\gamma_\mu)}_{m}(\mu,\mu_0,\omega, \omega_0)$.
We fix $\omega=\omega_0 =0.5, 1.0, 1.5, \ldots, 4.0$ and let $\mu$ vary between
1 and 4 GeV.
We find a good agreement for intermediate values of $\mu$ and $\omega$, where both
perturbation theory and lattice artefacts are expected to be under control.
There is also a good agreement for small values of $\mu$ (within
our statistical and systematic uncertainties) where we would have expected
non-perturbative effects to be more visible.
We also find that out of the two projectors, perturbation theory and lattice results
agree best in the $\gamma_\mu$-scheme.
On the other hand, the lattice artefacts for large values of $\mu$ and $\omega$
become relevant for $q^2 \gtrsim 25 \mathrm{GeV}^2$.
This becomes particularly visible for large values of $\omega = 4$, where
perturbation theory also becomes less reliable.

The only significant (relative) discrepancy we found is for $Z_q^{(\qslash)}$,
the quark wave function in the $\qslash$-scheme. However, this quantity
should be $\omega$-independent (up to lattice artefacts) 
and has no $\mu$-dependence at leading order (in the Landau gauge).
We show our results in Tables~\ref{table:sigmaq_NNNLO} and
\ref{table:sigmaq_conv_NNNLO}.
In this case the perturbative prediction is known at N$^3$LO.
As we can see from these tables, the series converges very poorly
in the sense that the relative difference
decreases very slowly as we increase the order of the expansion.
The difference between the non-perturbative 
result and the N$^3$LO prediction, namely $\sim 1.0195 - 1.0113 \sim 0.0082$,
could then be explained by higher corrections.
On the other hand, for $X=\gamma_\mu$, we find a much better convergence
of the perturbative expansion and a good agreement between the perturbative
and non-perturbative running after conversion to $\MSbar$.
\\


\begin{table}[t]
  \begin{center}
    \begin{tabular} {|c| c         c        c       c        c      |}
      \hline
      Scheme      & LO        & NLO       & NNLO      & NNNLO     & NP    \\
      \hline
      $\MSbar$     &  1.0      &  1.0048    &  1.0062  &  1.0064   &       \\
      $\MSbar\leftarrow \gamma_\mu $ &  1.0   &   1.0069  & 1.0078 & N.A. & \\
      $\MSbar\leftarrow \qslash $ &  1.0      &  1.0195   &  1.0175   &  1.0146 &  \\   
      $\gamma_\mu$  & 1.0       &  1.0017    &  1.0020  &  N.A      &  1.0037(20)\\
      $\qslash$    &  1.0      &  1.0048    &  1.0081  &  1.0113   &  1.0195(25) \\
      \hline
    \end{tabular}
  \end{center}
  \caption{Running between 2 and 2.5 GeV for the quark wave function in $\MSbar$
    and in the SMOM schemes $\gamma_\mu$($\omega=1$)  and  $\qslash$.
    In this case the running is known at NNNLO. }
\label{table:sigmaq_NNNLO}
\end{table}

\begin{table}[t]
\begin{center}
  \begin{tabular} { |c| c         c        c       |}
    \hline
    Scheme       &  NLO-LO  &  NNLO-NLO  &  NNNLO-NNLO   \\
    \hline
   $\MSbar$     &   0.0048 &  0.0013   & 0.0003  \\
   $\gamma_\mu$  &   0.0017 &  0.0003   &         \\
    $\qslash$   &   0.0048 &  0.0033   & 0.0032  \\
    \hline
\end{tabular}
  \end{center}
\caption{Study of the convergence of the perturbative series for running of the quark wave function
  between 2 and 2.5 GeV in $\MSbar$,  SMOM-$\gamma_\mu$  and $\qslash$. }
\label{table:sigmaq_conv_NNNLO}
\end{table}

\section{Conclusions and outlook}
We have implemented several IMOM schemes defined via two different projectors
and determined the renormalisation factors and
non-perturbative scale evolution functions of the quark mass and wave function.
We find that the non-pertubative and perturbative results agree very
well as long as we stay from the corner of the $\omega,\mu$ plane,
with one exception, namely $Z_q^{(\qslash)}$. There, we argued that
the reason for this relatively bad agreement is the poor convergence
of the perturbative expansion. We have shown some cases where $\omega\sim 2.0 - 2.5$
lead to substantially reduced infrared contamination and better control over
the discretisation effects, compared to standard SMOM kinematics.\\

We used two lattice spacings in this proof of concept study, 
clearly adding a finer lattice could potentially allow us to
probe the Rome-Southampton window even further.
It will also be interesting to extend this study to the case of
four-quark operators where the infrared contaminations due to chiral symmetry
breaking are significantly more sizeable.
The hope is that increasing the value of $\omega$ will reduce these contaminations
(compared to $\omega=1$) as it does for the bilinears.\\

\section{Acknowledgements} \label{ackn}
This work was supported by the Consolidated Grant ST/T000988/1 and
the work of JAG by a DFG Mercator Fellowship. The quark propagators
were computed on the DiRAC Blue Gene Q Shared Petaflop system at
the University of Edinburgh, operated by the Edinburgh Parallel
Computing Centre on behalf of the STFC DiRAC HPC Facility
(www.dirac.ac.uk). This equipment was funded by BIS National
E-infrastructure capital grant ST/K000411/1, STFC capital grant
ST/H008845/1, and STFC DiRAC Operations grants ST/K005804/1 and
ST/K005790/1. DiRAC is part of the National E-Infrastructure.  \\

We warmly thank our colleagues of the RBC and UKQCD collaborations.
We are particularly indebted to Peter Boyle,  Andreas J\"uttner,
J~Tobias Tsang for many interesting discussions. We also thank Peter
Boyle for his help with the UKQCD hadron software. We wish to thank
Holger Perlt for his early contribution in this area. N.G. thanks
his collaborators from the California Lattice (CalLat) collaboration
and in particular those working on NPR: David Brantley,  Henry
Monge-Camacho, Amy Nicholson and Andr\'e Walker-Loud. \\

\bibliography{biblio}
\bibliographystyle{utphys}


\end{document}